\documentstyle[amssymb,preprint,aps]{revtex}
\tightenlines 

\begin{document}
\title{Confined states in two-dimensional flat elliptic quantum dots and elliptic
quantum wires}
\author{M. van den Broek and F. M. Peeters\cite{email}}
\address{Departement Natuurkunde, Universiteit Antwerpen, Universiteitsplein 1, B-2610%
\\
Antwerpen, Belgium}
\date{23 March 2001}
\maketitle

\begin{abstract}
The energy spectrum and corresponding wave functions of a flat quantum dot
with elliptic symmetry are obtained exactly. A detailed study is made of the
effect of ellipticity on the energy levels and the corresponding wave
functions. The analytical behavior of the energy levels in certain limiting
cases is obtained.


\end{abstract}

\section{Introduction}

\bigskip Quantum dots have been the subject of both experimental and
theoretical research in recent years\cite{dots}. They have been successfully
created experimentally by applying lithographic and etching techniques to
impose a lateral structure onto an otherwise two-dimensional electron
system. These structures introduce electrostatic potentials in the plane of
the two-dimensional electron gas, which confine the electrons to a dot
region. The energy levels of electrons in such a quantum dot are fully
quantized like in an atom, and therefore they are also referred to as {\it %
artificial atoms}. For a theoretical description of such systems it is
essential to know the confinement potential. Often a parabolic potential is
used. Another simple model, which also has the advantage of representing a
finite confinement, is a piecewise constant potential with a circular
boundary\cite{peeters}. This type of potential represents more closely the
situation in self-assembled quantum dots\cite{dots}. The latter model is
used, but extended to an asymmetric situation, i.e. an elliptic dot.

There have been many calculations of energy levels in rectangular dots and
wires, see for example \cite{worlock,bryant,gershoni}, but in general, the
problem of even one particle confined in such a potential profile is not
exactly solvable, because the corresponding Schr\"{o}dinger equation is not
separable. Nevertheless one usually assumes separability, although this
approximation is only acceptable for deep wells, where only a negligible
part of the wave function is situated outside the dot. In our study we use
an elliptically shaped potential profile for which exact solutions can be
found, even when we take $\left. \Psi \right| _{\text{interface}}\neq 0$, or
a potential $V_{0}\ll \infty $ outside the dot. We are especially interested
in the results for elongated dots, which could then be compared with the
case of a rectangular slab or even a quantum wire. Our results are also
applicable to elliptic quantum wires.

In the literature one can find many references to calculations on elliptic
problems, we found one particularly useful\cite{morse}, in which the
diffraction of waves by ribbons and by slits is studied exactly. The problem
of a three-dimensional ellipsoidal (3D) quantum well has been solved exactly
in Refs. \cite{granger} and \cite{cantele} for $\left. \Psi \right| _{\text{%
interface}}=0$. To our knowledge, the corresponding 2D problem for finite
potential barriers has not been solved.

The paper is organized as follows. In Sec. II, we obtain the analytical
solution for the energy and the wave functions. The numerical results are
presented in Sec. III. Our conclusions are given in Sec. IV. A
representation \ for the Mathieu functions is listed in the appendix.

\section{The problem and equations}

A particle with mass $m$ is confined in a two-dimensional hard-wall
potential well of finite height with elliptic shape as shown in Fig. \ref
{fig1}. The potential profile we consider is constant everywhere, $V(x,y)=0$
inside the dot (i.e. $x^{2}/a^{2}+y^{2}/b^{2}<1$) and $V(x,y)=V_{0}$ outside
the dot. Stationary energy levels and wave functions are found by solving
the time-independent Schr\"{o}dinger equation

\begin{equation}
-\frac{\hbar ^{2}}{2m}{\bf \nabla }^{2}\Psi (x,y)+V(x,y)\Psi (x,y)=E\Psi
(x,y).
\end{equation}
Transformation from Cartesian to elliptical coordinates $x=f\cosh u\cos v,$ $%
0\leq u\leq \infty ,$ $y=f\sinh u\sin v,$ $0\leq v\leq 2\pi $, where $f=%
\sqrt{a^{2}-b^{2}}$ is the focal length, enables us to write the boundary of
the ellipse in a one-variable equation $u=U=%
\mathop{\rm arctanh}%
{}(b/a)$. Introducing the electron momentum $k^{2}=\frac{2m}{\hbar ^{2}}%
(E-V) $ the equation to be solved becomes 
\begin{equation}
\frac{\partial ^{2}\Psi }{\partial u^{2}}+\frac{\partial ^{2}\Psi }{\partial
v^{2}}+\frac{f^{2}k^{2}}{2}\cosh 2u\;\Psi -\frac{f^{2}k^{2}}{2}\cos 2v\;\Psi
=0.
\end{equation}
Proposing a solution $\Psi (u,v)=F(u)G(v)$ we find that the problem is
separable in two equations in $u$ and $v$, coupled by a separation constant $%
c$ and the variable $k$. Denoting $q=f^{2}k^{2}/4$ we obtain, 
\begin{mathletters}
\begin{eqnarray}
\frac{d^{2}F(u)}{du^{2}}-(c-2q\cosh 2u)F(u) &=&0,  \label{Fdiffequation} \\
\frac{d^{2}G(v)}{dv^{2}}+(c-2q\cos 2v)G(v) &=&0,  \label{Gdiffequation}
\end{eqnarray}
which are the characteristic equations for the Mathieu functions. Physical
solutions to these equations are well-known\cite{abramowitz} 
\end{mathletters}
\begin{mathletters}
\begin{eqnarray}
G_{l}(v,q) &=&\left\{ 
\begin{array}{c}
\mathop{\rm ce}%
_{l}(v,q), \\ 
\mathop{\rm se}%
_{l}(v,q),
\end{array}
\right. \\
F_{l}^{\text{in}}(u,q) &=&\left\{ 
\begin{array}{c}
\mathop{\rm Mc}%
_{l}^{(1)}(u,q_{1}), \\ 
\mathop{\rm Ms}%
_{l}^{(1)}(u,q_{1}),
\end{array}
\right. \\
F_{l}^{\text{out}}(u,q) &=&\left\{ 
\begin{array}{c}
\mathop{\rm Mc}%
_{l}^{(3)}(u,-q_{2}), \\ 
\mathop{\rm Ms}%
_{l}^{(3)}(u,-q_{2}),
\end{array}
\right.
\end{eqnarray}
where $l=0,1,2,\ldots $.The normalization factors are discarded and will be
incorporated later in the total wave function $\Psi (u,v)$. The Mathieu
functions $%
\mathop{\rm ce}%
_{l}(v,q)$ and $%
\mathop{\rm se}%
_{l}(v,q)$ are respectively even and odd in $v$ and are the analogue of the
cosine and sine functions in the angular solution of the circular dot
problem. For the $u$-problem we make a distinction between solutions inside
and outside the dot. The former show oscillatory behavior and are the
analogue of the Bessel functions $J_{l}(\rho )$ in the radial solution of
the circular problem. The latter are monotonically decaying functions and
are the analogue of the modified Bessel functions $K_{l}(\rho )$ in the
circular problem. In choosing these particular functions we used the
boundary conditions that the wave functions need to be finite when $u=0$ and 
$u\rightarrow \infty $ for obvious normalization reasons. Outside the dot $%
E<V_{0}$ and the variable $q$ becomes negative, which we noted explicitly.
In the appendix explicit expressions for the Mathieu functions are given.

The wave function $\Psi (u,v)$ and its derivatives must be continuous
everywhere. This condition requires that at every point on the border $(U,v)$
of the quantum dot we have 
\end{mathletters}
\begin{equation}
\frac{\left. \partial F_{l}^{\text{in}}(u,q)/\partial u\right| _{u=U,q=q_{1}}%
}{F_{l}^{\text{in}}(U,q_{1})}=\frac{\left. \partial F_{l}^{\text{out}%
}(u,q)/\partial u\right| _{u=U,q=-q_{2}}}{F_{l}^{\text{out}}(U,-q_{2})},
\label{boundarycondition}
\end{equation}
where the $u$-independent functions $G_{l}(v,q)$ cancel out.

We introduce $\sigma =\sqrt{ab}$, which is related to the area of the
ellipse and define 
\begin{equation}
\xi =\sigma k_{1,}\quad \eta =\sigma k_{2,}\quad \gamma =\sigma \sqrt{\frac{%
2m}{\hbar ^{2}}V_{0}}\text{ ,}
\end{equation}
with $k_{1}^{2}=\frac{2m}{\hbar ^{2}}E$ and $k_{2}^{2}=\frac{2m}{\hbar ^{2}}%
(V_{0}-E)$. The energy levels will be expressed in units of $V_{0}$. Clearly 
$\xi ^{2}+\eta ^{2}=\gamma ^{2}$ and $q_{1}=\frac{f^{2}k_{1}^{2}}{4}=\frac{1%
}{4}\frac{f^{2}}{\sigma ^{2}}\xi ^{2}=\frac{1}{4}(\frac{a}{b}-\frac{b}{a}%
)\xi ^{2}$. We are therefore left with two equations for even and odd
solutions in one variable $\xi $ to be solved numerically 
\begin{mathletters}
\begin{eqnarray}
\frac{\left. \partial 
\mathop{\rm Mc}%
_{l}^{(1)}/\partial u\left( u,\frac{1}{4}(\frac{a}{b}-\frac{b}{a})\xi
^{2}\right) \right| _{u=U}}{%
\mathop{\rm Mc}%
_{l}^{(1)}\left( U,\frac{1}{4}(\frac{a}{b}-\frac{b}{a})\xi ^{2}\right) } &=&%
\frac{\left. \partial 
\mathop{\rm Mc}%
_{l}^{(3)}/\partial u\left( u,-\frac{1}{4}(\frac{a}{b}-\frac{b}{a})(\gamma
^{2}-\xi ^{2})\right) \right| _{u=U}}{%
\mathop{\rm Mc}%
_{l}^{(3)}\left( U,-\frac{1}{4}(\frac{a}{b}-\frac{b}{a})(\gamma ^{2}-\xi
^{2})\right) }, \\
\frac{\left. \partial 
\mathop{\rm Ms}%
_{l}^{(1)}/\partial u\left( u,\frac{1}{4}(\frac{a}{b}-\frac{b}{a})\xi
^{2}\right) \right| _{u=U}}{%
\mathop{\rm Ms}%
_{l}^{(1)}\left( U,\frac{1}{4}(\frac{a}{b}-\frac{b}{a})\xi ^{2}\right) } &=&%
\frac{\left. \partial 
\mathop{\rm Ms}%
_{l}^{(3)}/\partial u\left( u,-\frac{1}{4}(\frac{a}{b}-\frac{b}{a})(\gamma
^{2}-\xi ^{2})\right) \right| _{u=U}}{%
\mathop{\rm Ms}%
_{l}^{(3)}\left( U,-\frac{1}{4}(\frac{a}{b}-\frac{b}{a})(\gamma ^{2}-\xi
^{2})\right) }.
\end{eqnarray}
Solutions are indexed $\xi _{n,l},n=1,2,\ldots ;l=0,1,\ldots $. The energy
spectrum is then given by 
\end{mathletters}
\begin{equation}
E_{n,l}=\frac{\hbar ^{2}}{2m}k_{1_{n,l}}^{2}=\frac{\xi _{n,l}^{2}}{\gamma
^{2}}V_{0}.
\end{equation}

In the limit of an infinitely high barrier, $V_{0}\rightarrow \infty $, the
boundary condition will simply become $F_{l}^{\text{in}}(u=U,q_{1})=0$, or
equivalently, 
\begin{mathletters}
\begin{eqnarray}
\mathop{\rm Mc}%
\nolimits_{l}^{(1)}\left( U,\frac{1}{4}(\frac{a}{b}-\frac{b}{a})\xi
^{2}\right) &=&0, \\
\mathop{\rm Ms}%
\nolimits_{l}^{(1)}\left( U,\frac{1}{4}(\frac{a}{b}-\frac{b}{a})\xi
^{2}\right) &=&0.
\end{eqnarray}

When the particle has a different effective mass inside ($m_{1}$) and
outside the dot ($m_{2}$) the boundary condition Eq. (\ref{boundarycondition}%
) has to be modified into 
\end{mathletters}
\begin{equation}
\frac{\left. \partial F_{l}^{\text{in}}/\partial u(u,q_{1})\right| _{u=U}}{%
m_{1}F_{l}^{\text{in}}(U,q_{1})}=\frac{\left. \partial F_{l}^{\text{out}%
}/\partial u(u,-q_{2})\right| _{u=U}}{m_{2}F_{l}^{\text{out}}(U,-q_{2})}%
\text{ .}
\end{equation}

The explicit expressions for the wave functions are, for the even functions, 
\begin{equation}
\Psi _{n,l}^{e}(u,v)=\left\{ 
\begin{array}{c}
N_{n,l}^{e}%
\mathop{\rm Mc}%
_{l}^{(1)}(u,q_{1_{n,l}})%
\mathop{\rm ce}%
_{l}(v,q_{1_{n,l}}),\qquad u\leq U, \\ 
N_{n,l}^{e}%
\mathop{\rm Mc}%
_{l}^{(3)}(u,-q_{2_{n,l}})%
\mathop{\rm ce}%
_{l}(v,-q_{2_{n,l}}),\qquad u>U,
\end{array}
\right.
\end{equation}
and for the odd functions, 
\begin{equation}
\Psi _{n,l}^{o}(u,v)=\left\{ 
\begin{array}{c}
N_{n,l}^{o}%
\mathop{\rm Ms}%
_{l}^{(1)}(u,q_{1_{n,l}})%
\mathop{\rm se}%
_{l}(v,q_{1_{n,l}}),\qquad u\leq U, \\ 
N_{n,l}^{o}%
\mathop{\rm Ms}%
_{l}^{(3)}(u,-q_{2_{n,l}})%
\mathop{\rm se}%
_{l}(v,-q_{2_{n,l}}),\qquad u>U,
\end{array}
\right.
\end{equation}
where $N_{n,l}^{e}$ and $N_{n,l}^{o}$ are determined by normalization of the
complete wave function.

\section{Numerical results}

In Fig. \ref{fig2} we show the dependence of the energy of the bound states
on the strength of the well $\gamma =\sqrt{\frac{2m}{\hbar ^{2}}abV_{0}}$,
which depends both on the potential height $V_{0}$ and on the area of the
dot $ab$. For clarity we have limited all our graphs to $l\leq 7$. For $%
a/b=1 $ the results coincide with our calculations for a circular dot. When
the ratio $a/b$ increases, the degeneracy of the $l\neq 0$ energy levels is
lifted and they split in separate levels. This is a result of a decrease in
symmetry of the system in the transition from a circular to an elliptic dot.
With increasing $a/b$,\ energy levels corresponding with even states lower,
while those corresponding with uneven states rise in energy. This behavior
is more clearly visible in Fig. \ref{fig3}, where we have plotted the energy
levels against the ratio $a/b$, with $\gamma $ fixed. For small and narrow
dots, only the even $n=1$ states remain, while the uneven states become
unbound. The total number of states stays approximately equal. In Fig. \ref
{fig3} it is also clear that even the smallest and narrowest dot holds at
least one confined state. The energy of this lowest state is shown in Fig. 
\ref{fig4} as a function of the eccentricity $a/b$. There is only a small
increase of the energy level with increasing eccentricity.

The asymptotic behavior for the energy of the ground state in a shallow 2D
well can be obtained analytically. Because of the weak dependence of this
energy on the eccentricity we assume for simplicity $a/b=1$. Outside the
well the wavefunction is then given by the modified Bessel function, 
\begin{equation}
\Psi \left( \rho \right) =K_{0}\left( \sqrt{\frac{2m}{\hbar ^{2}}\left|
E_{0}\right| }\rho \right) \text{,}
\end{equation}
where $\left| E_{0}\right| =V_{0}-E$ is the binding energy in the well and $%
\rho =\left( x^{2}+y^{2}\right) ^{1/2}$. For small $\left| E_{0}\right| $
this wave function can to first order be approximated by 
\begin{equation}
\Psi \left( \rho \right) \thickapprox -\ln \left( \frac{e^{g}}{2}\sqrt{\frac{%
2m}{\hbar ^{2}}\left| E_{0}\right| }\rho \right) \text{,}
\end{equation}
with $g=0.57721\ldots $ the constant of Euler-Mascheroni. Inside the
potential well the ground state wave function is almost constant.
Integrating this function over the Schr\"{o}dinger equation we obtain the
binding energy 
\begin{equation}
\left| E_{0}\right| =\frac{4}{e^{2g}}\frac{\hbar ^{2}}{2ma^{2}}\exp \left( -%
\frac{2\hbar ^{2}}{mV_{0}a^{2}}\right)
\end{equation}
with the prefactor $4/e^{2g}=1.26096\ldots $. In Figs. \ref{fig2} and \ref
{fig3} the ground state energy for small $\gamma $ can be approximated by 
\begin{equation}
E/V_{0}=1-\frac{1.26}{\gamma ^{2}}e^{-4/\gamma ^{2}}\text{.}
\end{equation}

The results for an infinitely high confinement potential are shown in Fig. 
\ref{fig5}. The behavior is comparable to the finite well case, except that
all the energy levels will eventually rise with increasing eccentricity.

Notice that in Fig. \ref{fig5} we find that for large eccentricity the
energy levels appear in bands where the levels are practically equidistant.
This can be understood from the following simple consideration. If $b/a\ll 1$
we can use an adiabatic approximation and assume that the motion along the $%
y $-direction is much faster than along $x$. For $\left| x\right| <a$ the
electron feels a hard wall potential along the $y$-direction of width $W=2b%
\sqrt{1-x^{2}/a^{2}}$ which has energy levels $E_{y}=\frac{\pi ^{2}\hbar ^{2}%
}{2mW}n_{y}^{2}$. This results into a potential along the $x$-direction $%
V(x)=\frac{\pi ^{2}\hbar ^{2}}{8mb^{2}}\frac{n_{y}^{2}}{(1-x^{2}/a^{2})}%
\simeq \frac{\pi ^{2}\hbar ^{2}}{8mb^{2}}n_{y}^{2}(1+x^{2}/a^{2})$ which
near its minimum is parabolic. The resulting energy levels become $%
E_{n_{x},n_{y}}=\frac{\pi ^{2}\hbar ^{2}}{8mb^{2}}n_{y}^{2}+\frac{\pi
^{2}\hbar ^{2}}{2mab}n_{y}\left( n_{x}+1/2\right) $, which for fixed $n_{y}$
results into an equidistant set of levels as found in Fig. \ref{fig5} for $%
a/b\gg 1$. The dependence on the eccentricity becomes more clear if we
introduce $\sigma =\sqrt{ab}$, which was taken fixed in Fig. \ref{fig5},
into the above expression, which leads to 
\begin{equation}
E_{n_{x},n_{y}}=\frac{\pi ^{2}\hbar ^{2}}{8m\sigma ^{2}}n_{y}^{2}\frac{a}{b}+%
\frac{\pi ^{2}\hbar ^{2}}{2m\sigma ^{2}}n_{y}\left( n_{x}+1/2\right) \text{.}
\end{equation}
Notice that the linear increase with $a/b$ results from the zero point
motion along the $y$-direction.

To see the effect of different well ($m_{1}$) and barrier ($m_{2}$) masses
we present in Fig. \ref{fig6} the results for the bound states in the case
of three different mass ratios $\mu =m_{2}/m_{1}=0.5,1$ and $2$. There is a
decrease of the energy levels, i.e. increase of binding, when the effective
mass of the electron outside the dot is higher than inside the dot. The $l=0$
levels become bound at the same value for $\gamma $, while this is no longer
the case for $l>0$.

The density distribution of the wave functions corresponding to the
different bound states, denoted by $(n,l)^{e}$ or $(n,l)^{o}$, are shown in
Fig. \ref{fig7} for $a/b=5$ and $\gamma =11$. The even $n=1$ states form the
lowest levels and they have all their extrema of the wave function located
along a row in the direction of the longest axis of the ellipse. In this
particular case, a second row of extrema appears only at the sixth state $%
(1,1)^{o}$, which is odd. When $n=2$, we see three rows appear. In all cases
the number of extrema in a row is determined by $l$. The general pattern is
that the lowest levels are those for which the position of the extrema of
the wave function best fit the shape of the quantum dot.

\section{Conclusions}

By expressing the wave functions in the appropriate coordinate system, viz.
elliptical coordinates, we were able to find analytical solutions for the
Schr\"{o}dinger equation describing an elliptical quantum dot with finite
height hard walls. These functions are the Mathieu functions. The condition
of smoothness of the wave function at the boundary of the dot, results in an
expression for the energy levels. The main effect of the elongation of the
dot from a circular to an elliptic shape is a lifting of the degeneracy of
the $l>1$ levels. We found exact results for the energy levels and the wave
functions for arbitrary quantum numbers. The basic solutions found here may
be the basis for more advanced problems with elliptic geometry, e.g. many
particle quantum dots and quantum dots in a magnetic field. The results can
readily be expanded to the case of an elliptic wire. The particle can then
move freely along the $z$-direction of a wire with an elliptic
cross-section. For the energy states we find 
\begin{equation}
\widetilde{E}_{n,l,k_{z}}=\frac{\hbar k_{z}^{2}}{2m}+E_{n,l}\text{,}
\end{equation}
where $E_{n,l}$ are the above discussed levels\ and the wave functions are 
\begin{equation}
\widetilde{\Psi }(x,y,z)=e^{ik_{z}z}\Psi _{n,l}(x,y)\text{.}
\end{equation}

\section*{Acknowledgments}

This work was supported by the Flemish Science Foundation (FWO-Vl), IUAP-IV,
and the ''Bijzonder Onderzoeksfonds van de Universiteit Antwerpen''. F. M.
P. acknowledges discussions with J. M. Worlock in the initial stage of this
work.

\section*{Appendix}

Solutions to Eq. (\ref{Gdiffequation}) are found in the form of a Fourier
series. Physically relevant solutions obey periodic boundary conditions 
\begin{mathletters}
\begin{eqnarray}
G(v) &=&G(v+\pi ),  \eqnum{A.1a} \\
G(v) &=&G(v+2\pi ).  \eqnum{A.1b}
\end{eqnarray}
Even periodic solutions are expanded in a Fourier series of cosine
functions, while odd periodic solutions are expanded in a Fourier series of
sine functions. In this way, we find the four fundamental solutions, which
are called the Mathieu functions\cite{abramowitz,zhang,methods}

\end{mathletters}
\begin{eqnarray}
\mathop{\rm ce}%
\nolimits_{2l}(v,q) &=&\sum_{j=0}^{\infty }A_{2j}^{2l}(q)\cos 2jv, 
\eqnum{A.2a} \\
\mathop{\rm ce}%
\nolimits_{2l+1}(v,q) &=&\sum_{j=0}^{\infty }A_{2j+1}^{2l+1}(q)\cos (2j+1)v,
\eqnum{A.2b} \\
\mathop{\rm se}%
\nolimits_{2l+1}(v,q) &=&\sum_{j=0}^{\infty }B_{2j+1}^{2l+1}(q)\sin (2j+1)v,
\eqnum{A.2c} \\
\mathop{\rm se}%
\nolimits_{2l+2}(v,q) &=&\sum_{j=0}^{\infty }B_{2j+2}^{2l+2}(q)\sin (2j+2)v.
\eqnum{A.2d}
\end{eqnarray}
By substituting\ these series into Eq. (\ref{Gdiffequation}), we find a set
of recursion relations for the expansion coefficients, e.g. for $%
A_{2j}^{2l}(q)$%
\begin{eqnarray}
cA_{0}^{2l}-qA_{2}^{2l} &=&0,  \nonumber \\
(c-4)A_{2}^{2l}-q(2A_{0}^{2l}+A_{4}^{2l}) &=&0,  \eqnum{A.3} \\
\left[ c-(2l)^{2}\right] A_{2j}^{2l}-q(A_{2j-2}^{2l}+A_{2j+2}^{2l})
&=&0\quad (j\geq 2).  \nonumber
\end{eqnarray}

Numerical methods to calculate the Mathieu functions are described in detail
in the literature\cite{alhargan,leeb,shirts,shirts2,lindner,zhang}.
Solutions to Eq. (\ref{Fdiffequation}) are written as an expansion in Bessel
functions. With $u_{1}=\sqrt{q}e^{-u}$ and $u_{2}=\sqrt{q}e^{u}$\ we find 
\begin{eqnarray}
\mathop{\rm Mc}%
\nolimits_{2l}^{(1)}(u,q) &=&\frac{1}{A_{0}^{2l}(q)}\sum_{j=0}^{\infty
}(-1)^{j+l}A_{2j}^{2l}(q)J_{j}(u_{1})J_{j}(u_{2}),  \eqnum{A.4a} \\
\mathop{\rm Mc}%
\nolimits_{2l+1}^{(1)}(u,q) &=&\frac{1}{A_{1}^{2l+1}(q)}\sum_{j=0}^{\infty
}(-1)^{j+l}A_{2j+1}^{2l+1}(q)\left[
J_{j}(u_{1})J_{j+1}(u_{2})+J_{j+1}(u_{1})J_{j}(u_{2})\right] ,  \eqnum{A.4b}
\\
\mathop{\rm Ms}%
\nolimits_{2l+1}^{(1)}(u,q) &=&\frac{1}{B_{1}^{2l+1}(q)}\sum_{j=0}^{\infty
}(-1)^{j+l}B_{2j+1}^{2l+1}(q)\left[
J_{j}(u_{1})J_{j+1}(u_{2})-J_{j+1}(u_{1})J_{j}(u_{2})\right] ,  \eqnum{A.4c}
\\
\mathop{\rm Ms}%
\nolimits_{2l+2}^{(1)}(u,q) &=&\frac{1}{B_{2}^{2l+2}(q)}\sum_{j=0}^{\infty
}(-1)^{j+l}B_{2j+2}^{2l+2}(q)\left[
J_{j}(u_{1})J_{j+2}(u_{2})-J_{j+2}(u_{1})J_{j}(u_{2})\right] ,  \eqnum{A.4d}
\end{eqnarray}
and 
\begin{eqnarray}
\mathop{\rm Mc}%
\nolimits_{2l}^{(3)}(u,-q) &=&\frac{2i}{\pi }\frac{(-1)^{l+1}}{A_{0}^{2l}(q)}%
\sum_{j=0}^{\infty }A_{2j}^{2l}(q)I_{j}(u_{1})K_{j}(u_{2}),  \eqnum{A.5a} \\
\mathop{\rm Mc}%
\nolimits_{2l+1}^{(3)}(u,-q) &=&\frac{2}{\pi }\frac{(-1)^{l+1}}{%
B_{1}^{2l+1}(q)}\sum_{j=0}^{\infty }B_{2j+1}^{2l+1}(q)\left[
I_{j}(u_{1})K_{j+1}(u_{2})-I_{j+1}(u_{1})K_{j}(u_{2})\right] ,  \eqnum{A.5b}
\\
\mathop{\rm Ms}%
\nolimits_{2l+1}^{(3)}(u,-q) &=&\frac{2}{\pi }\frac{(-1)^{l+1}}{%
A_{1}^{2l+1}(q)}\sum_{j=0}^{\infty }A_{2j+1}^{2l+1}(q)\left[
I_{j}(u_{1})K_{j+1}(u_{2})+I_{j+1}(u_{1})K_{j}(u_{2})\right] ,  \eqnum{A.5c}
\\
\mathop{\rm Ms}%
\nolimits_{2l+2}^{(3)}(u,-q) &=&\frac{2i}{\pi }\frac{(-1)^{l}}{%
B_{2}^{2l+2}(q)}\sum_{j=0}^{\infty }B_{2j+2}^{2l+2}(q)\left[
I_{j}(u_{1})K_{j+2}(u_{2})-I_{j+2}(u_{1})K_{j}(u_{2})\right] .  \eqnum{A.5d}
\end{eqnarray}
To obtain the derivatives of the Mathieu functions we used these series
expansions of the Mathieu functions and applied the analytical formulas for
the derivatives of Bessel functions\cite{abramowitz}.

\begin{figure}[tbp]
\caption{The potential profile $V(x,y)$ is zero inside the elliptic dot and
equal to $V_{0}$ outside the dot. The long axis of the ellipse is $2a$, the
short axis is $2b$.}
\label{fig1}
\end{figure}
\begin{figure}[tbp]
\caption{The confined even and odd states $(n,l)^{e/o\text{ }}$of an
elliptic quantum dot with eccentricity $a/b$ as a function of the strength
of the well $\protect\gamma $. All dots have at least one bound state. (a)
For $a/b=1$ the even and odd states are degenerate. (b) When $a\neq b$ this
degeneracy is lifted for $l>0$.}
\label{fig2}
\end{figure}
\begin{figure}[tbp]
\caption{The confined energy states $(n,l)^{e/o}$ of an elliptic quantum dot
with strength $\protect\gamma =9$ as a function of the eccentricity $a/b$.
All the even $n=1$ states decrease with increasing $a/b$.}
\label{fig3}
\end{figure}
\begin{figure}[tbp]
\caption{The energy of the ground state $(1,0)^{e}$ of shallow elliptic
quantum dots with small strengths $\protect\gamma =1,1.5,2,3$ and $4$ as a
function of the eccentricity $a/b$.}
\label{fig4}
\end{figure}
\begin{figure}[tbp]
\caption{The energy states in an infinite elliptic quantum well as a
function of the eccentricity $a/b$. The effective electron mass was taken to
be $m^{\ast }=0.041$ $m_{e}$ and the area of the dot $\protect\sigma =\left(
ab\right) ^{1/2}=10$ nm.}
\label{fig5}
\end{figure}
\begin{figure}[tbp]
\caption{The confined energy states of an elliptic quantum dot with
eccentricity $a/b=5$ as a function of the strength of the well $\protect%
\gamma $ for different mass ratio $\protect\mu =m_{2}/m_{1}$ where $m_{1}$ ($%
m_{2}$) is the effective mass of the electron inside (outside) the dot. For
clarity we have split up the graph for different $l$.}
\label{fig6}
\end{figure}
\begin{figure}[tbp]
\caption{The probablity density for the eigenfunctions $(n,l)^{e/o}$ of all
the bound states in an elliptic quantum dot with strength $\protect\gamma =11
$ and ratio $a/b=5$ in order of increasing energy. The lowest states have $%
n=1$ and are all even: the maxima in the probability are along the long axis
of the ellipse. Only higher energy states can have more rows of maxima.}
\label{fig7}
\end{figure}

\end{document}